\documentclass[aps,prm,twocolumn,floatfix,10pt]{revtex4}
\usepackage{graphicx}
\usepackage{amsmath}
\usepackage{color}
\usepackage{wrapfig}
\usepackage{latexsym}
\usepackage[capitalize]{cleveref}
\usepackage{appendix}

\begin{document}

\title{Computing the thermal transport coefficient of neutral amorphous polymers using exact vibrational density of states: Comparison with experiments}

\author{Debashish Mukherji}
\email[]{debashish.mukherji@ubc.ca}
\affiliation{Quantum Matter Institute, University of British Columbia, Vancouver V6T 1Z4, Canada}


\begin{abstract}
Thermal transport coefficient $\kappa$ is an important property that often dictates broad applications of a polymeric material, 
while at the same time its computation remains challenging. In particular, classical simulations 
overestimate $\kappa$ than the experimentally measured $\kappa^{\rm exp}$ and thus hinder their meaningful comparison. 
This is even when very careful simulations are performed using the most accurate empirical potentials.
A key reason for such a discrepancy is because polymers have quantum--mechanical, nuclear degrees--of--freedom 
whose contribution to the heat balance is non--trivial. 
In this work, two semi--analytical approaches are considered to accurately compute $\kappa$ by 
using the exact vibrational density of states $g(\nu)$. The first approach is based within the framework 
of the minimum thermal conductivity model, while the second uses computed quantum heat capacity to scale $\kappa$. 
Computed $\kappa$ of a set of commodity polymers compares quantitatively with $\kappa^{\rm exp}$.
\end{abstract}

\maketitle

\section{Introduction}
\label{sec:intro}

Thermal transport coefficient $\kappa$ measures the ability of a material to conduct the heat current~\cite{CahillRev03JAP,RevModPhy07,Keblinski20,PolRevTT14,DMKKpolRev23}.
Here, $\kappa$ is directly related to the heat capacity $c$, the group velocity $v_{\rm g}$, 
and the phonon mean--free path $\Lambda = \tau v_{\rm i}$, with $\tau$ being the phonon life time~\cite{Kappabook}. 
Traditionally, extensive efforts have been devoted in investigating $\kappa$ behavior in the crystalline materials~\cite{RevModPhy07,donadio2009atomistic,lee2016thermal,prasher2009turning},
the recent interest is more devoted to the polymeric solids~\cite{Keblinski20,PolRevTT14,Pipe15NMat,Cahill16Mac,Cahill21SemiCrys,MM21acsn}. 
This is particularly because polymers are an important class of soft matter, where the relevant energy scale is of the order of 
$k_{\rm B}T$ at a temperature $T = 300$ K and $k_{\rm B}$ being the Boltzmann constant, and thus their properties are dictated by large conformational and 
compositional fluctuations~\cite{DGbook,DoiBook,Mueller20PPS,Mukherji20AR}.
This soft nature of polymers makes them important in designing flexible advanced materials with tunable thermal properties.

Polymers are a special case, where there are two main microscopic interactions, i.e., the intra--molecular interactions 
along a chain contour and the non--bonded interactions between the neighbouring monomers. 
In this context, $\kappa$ of amorphous polymers is dictated by the localized vibrations that 
are usually only within the range of direct non--bonded contacts (i.e., $\Lambda$ is very small) and thus are 
dominated by the monomer--monomer interactions~\cite{keb09jap,pmmalocalized},
which in the non--conducting polymers can either be van der Waals (vdW) or hydrogen bonded (H--bond)~\cite{Mukherji20AR,desiraju02}.
Because of the above reasons, polymers fall in the low $\kappa$ materials~\cite{Keblinski20,PolRevTT14},
having typical values that are several orders of magnitude smaller than the standard crystals~\cite{CahillRev03JAP,RevModPhy07}. 
For example, the experimentally measured $\kappa^{\rm exp} \simeq 0.1-0.2$ W/Km in vdW polymers ~\cite{Cahill16Mac,Mukherji19PRM}, 
while $\kappa^{\rm exp} \to 0.4$ W/Km in the H--bonded systems~\cite{Pipe15NMat,Cahill16Mac}. 

Extensive experimental and simulation efforts have been devoted to establish structure--property relationship in 
polymeric solids with a goal to obtain a tunable $\kappa$. Here, the standard classical simulation techniques 
are of particular importance. However, routinely employed classical setups often overestimate $\kappa^{\rm cl}$ in comparison to 
$\kappa^{\rm exp}$~\cite{KappaMDExp,martin21prm,kappaOil} and thus
hinders their meaningful comparison. Complexities get even more elevated when dealing with systems at 
different thermodynamic state points~\cite{plakappa,PMMACpexp}, complex macro--molecular architectures~\cite{cahill21acsappm,DMPRM21}, 
and/or relative compositions in the case of multi--component mixtures~\cite{Pipe15NMat,Cahill16Mac,Bruns19mac}.

One can simply argue that $\kappa^{\rm cl} > \kappa^{\rm exp}$ 
might be due to the inaccuracies in classical force--field parameters and in the $\kappa^{\rm cl}$ calculations. 
While simulation errors are certainly inevitable, it may still be presumptuous to come to such a 
trivial conclusion because of the complexities of underlying macro--molecular systems.
A closer look in an amorphous polymer reveals that the non--bonded interactions are soft that dictate polymer properties
(i.e., low $\nu$ anharmonic classical modes), while the intra--molecular interactions along a chain backbone are stiff~\cite{Tim19acsml,martin21prm,baschnagel22jcp}. 
For example, the vibrational frequency of a C--H bond in polymers is $\nu \simeq 90$ THz~\cite{Tim19acsml,martin21prm}. Note that C--H is a common building block of most commodity polymers.
Such a stiff mode and many other modes in a polymer remain quantum mechanically frozen at $T = 300$ K (with a representative frequency $\nu_{\rm room} \simeq 6.2$ THz).
On the contrary, however, a classical setup by default considers all modes (irrespective of their nature), 
thus overestimates $c$~\cite{BHOWMIK2019176,martin21prm} or $\kappa$~\cite{Mukherji19PRM,martin21prm,kappaOil} in polymers.

\begin{figure}[ptb]
\includegraphics[width=0.49\textwidth]{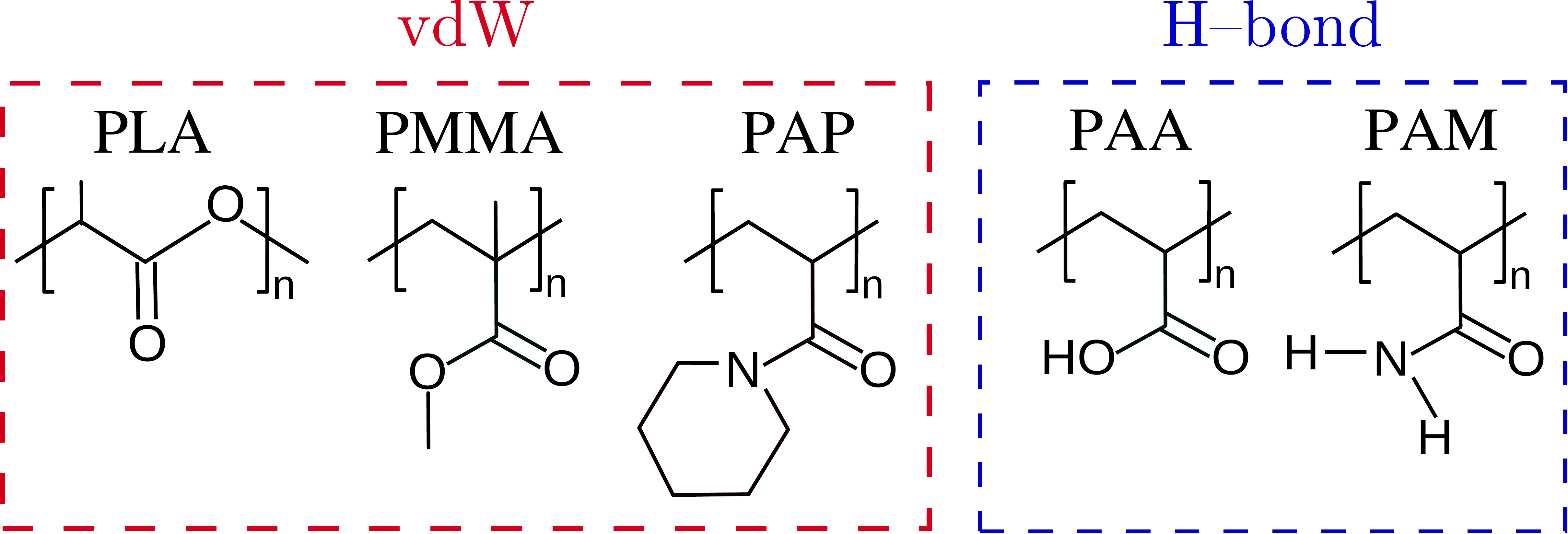}
\caption{Schematics representation of the commodity polymeric structures
investigated in this study. The red panel compiles the van der Waals (vdW) based systems, 
i.e., poly(lactic acid) (PLA), poly(methyl methacrylate)(PMMA), and poly(N-acryloyl piperidine) (PAP).
The blue panel shows two examples of hydrogen bonded (H--bond) polymers, 
i.e., poly(acrylic acid) (PAA) and polyacrylamide (PAM).}
\label{fig:struct}
\end{figure}

The discussions above pose a grand challenge on how to accurately compute $\kappa$ in polymeric solids
with a goal to achieve their meaningful (quantitative) comparison with $\kappa^{\rm exp}$.
Motivated by this need, the present work uses two simple semi--analytical approaches using the exact vibrational density of states $g(\nu)$
to estimate $\kappa$ at different thermodynamic state points. While the first approach (Approach I) is based within the well--known 
framework of the minimum thermal conductivity model (MTCM)~\cite{Cahill90PRB}, another approach (Approach II) estimates $\kappa$ 
by the accurate computation of quantum $c$~\cite{martin21prm}.
To validate our scenarios, this work investigates a set of experimentally relevant amorphous (commodity) polymers, see Fig.~\ref{fig:struct}.

Remainder of the draft is organized as follows: in Section~\ref{sec:mod} we highlight the system specific details 
investigated in this study and their force field parameters. The results are discussed in Section~\ref{sec:res}
and finally the conclusions are drawn in Section~\ref{sec:conc}.

\section{Materials, model, and method}
\label{sec:mod}

In this work, a set of five commodity polymers, covering across vdW and H--bonded systems: 
namely, poly(lactic acid) (PLA), poly(methyl methacrylate)(PMMA), poly(N-acryloylpiperidine) (PAP), polyacrylamide (PAM), and poly(acrylic acid) (PAA).
The monomer structures of these systems are shown in Fig.~\ref{fig:struct}.
The specific systems are chosen because their detailed experimental data are available~\cite{Cahill16Mac,plakappa,pmma14exp,pmmalocalized} and 
also because their available well--equilibrated configurations for all these samples~\cite{Mukherji19PRM,martin21prm}.

The chain length $N_{\ell} = 30$ is taken for all systems, except for PAM where $N_{\ell} = 32$. 
Each system consist of 200 chains within a cubic simulation box. 
The standard OPLS--AA force field parameters~\cite{OPLS} are used for PLA, PAP, and PAA, while a set of modified parameters are used 
for PMMA~\cite{Mukherji17NC} and PAM~\cite{Mukherji17JCP}. Simulations are performed using the GROMACS package~\cite{Abraham:2015}. 

Temperature is imposed using the velocity--rescaling thermostat~\cite{Vscale} with a damping time of $\tau_{T} = 1$ ps, 
and the pressure is set to 1 atm with a Berendsen barostat~\cite{Berend} with a time constant $\tau_{p} = 0.5$ ps. 
Electrostatics are treated using the particle--mesh Ewald method. The interaction cutoff for 
the non--bonded interactions are chosen as $r_{\rm c} = 1.0$ nm. 
The simulation time step is set to $\Delta t = 1$ fs during equilibration and the equations of motion are integrated using the leap--frog algorithm.

All these polymers were equilibrated earlier in their (solvent-free) melt states at $T = 600$ K for at--least 1 $\mu$s each sample, 
i.e., 500 ns in Ref.~\cite{Mukherji19PRM} and another 500 ns in Ref.~\cite{martin21prm}. Note that $T = 600$ K 
is at least 150 K above their calculated glass transition temperatures~\cite{Mukherji19PRM}. 
For this study, these melt equilibrated samples were individually quenched to $T=300$ K with a rate $0.04$ K/ns for a total of 7.5 $\mu$s per sample.
The total simulation time accumulated for this study alone is over 40 $\mu$s.

\section{Results and discussions}
\label{sec:res}

\subsection{Vibrational density of states}

\begin{figure}[h!]
\includegraphics[width=0.49\textwidth]{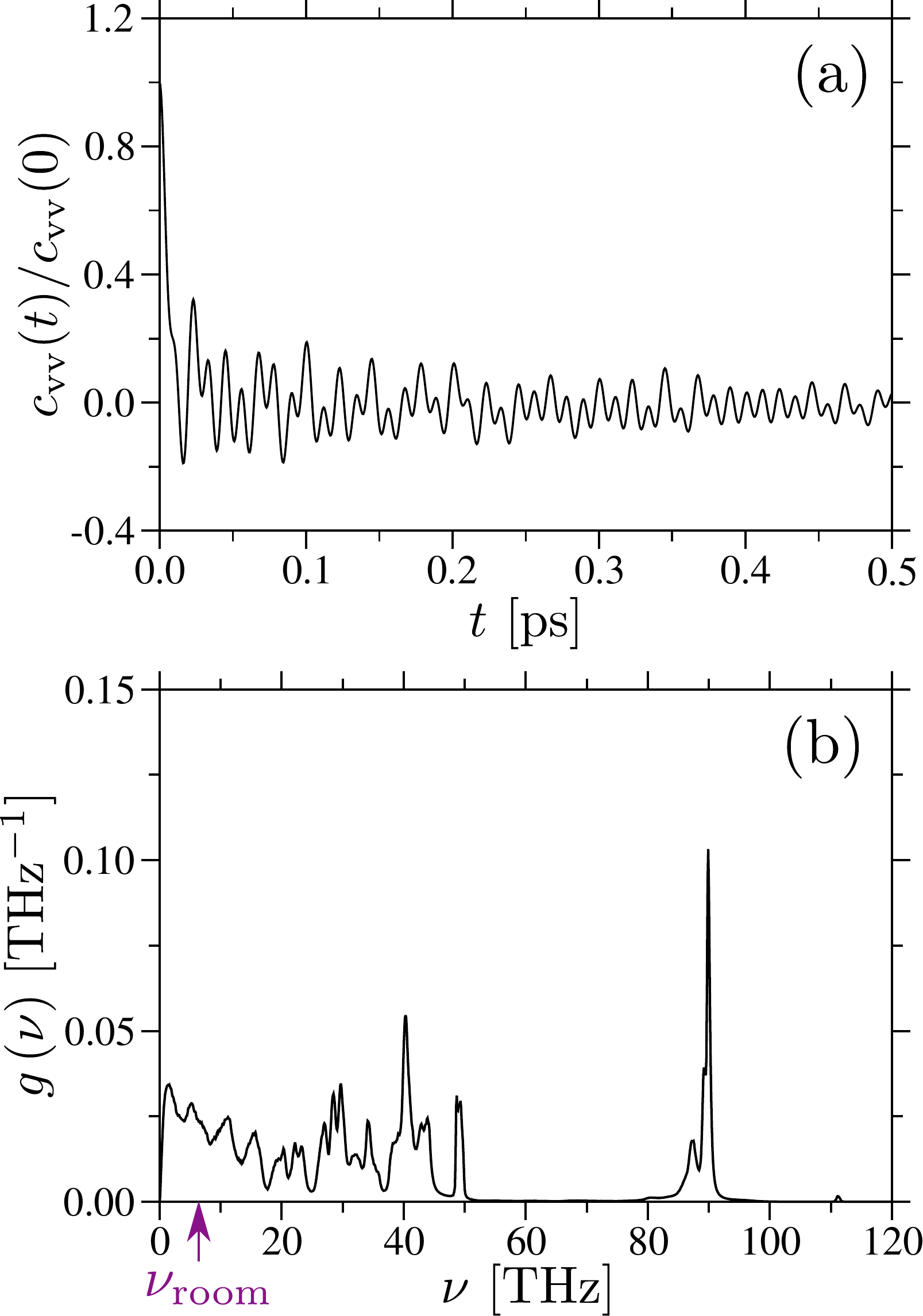}
\caption{Part (a) shows a normalized mass weighted velocity auto--correlation function $c_{\rm vv}(t)/c_{\rm vv}(0)$ 
for poly(lactic acid) (PLA) at a temperature $T = 300$ K. The
vibrational density of states $g(\nu)$ of PLA is shown in part (b), where
$g(\nu)$ is calculated using Eq.~\ref{eq:vdos}. The arrow in part (b) indicates at the vibrational 
frequency $\nu_{\rm room} \simeq 6.2$ THz corresponding to $T = 300$ K.}
\label{fig:vdosPLA}
\end{figure}

\begin{figure}[ptb]
\includegraphics[width=0.49\textwidth]{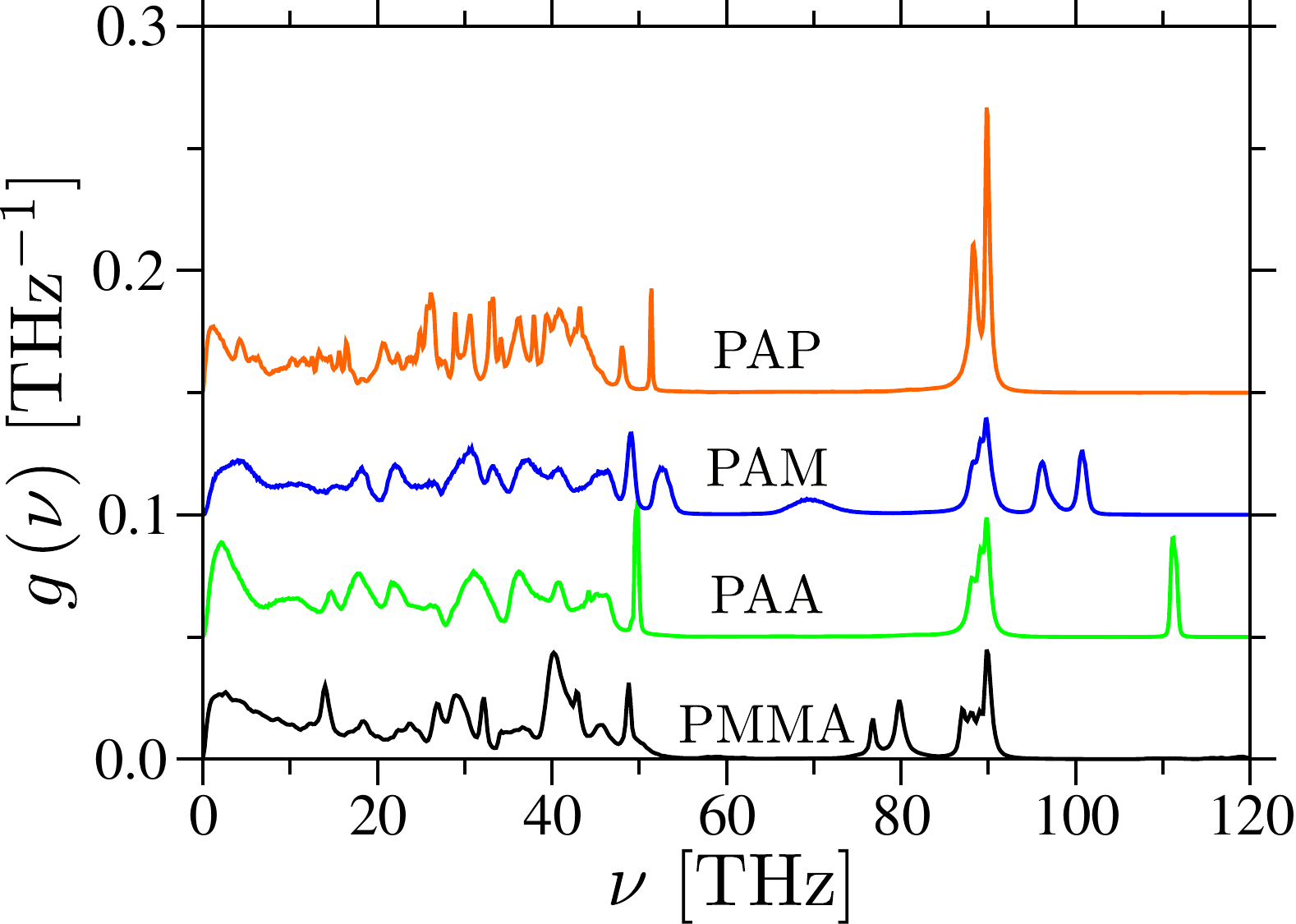}
\caption{Same as Fig.~\ref{fig:vdosPLA}(b), however, for four other commodity polymers, namely; poly(N-acryloyl piperidine) (PAP), polyacrylamide (PAM), poly(acrylic acid) (PAA), and poly(methyl methacrylate)(PMMA).
Individual $g(\nu)$ are shifted for a clearer representation.} 
\label{fig:gnu_4pol}
\end{figure}

A key observable for this study is $g(\nu)$. For this purpose, the mass--weighted velocity auto--correlation function
is calculated using,
\begin{equation}
    c_{\rm vv}(t) = \sum_{i} m_i \langle {\overrightarrow v}_i(t) \cdot {\overrightarrow v}_i(0)\rangle.
    \label{eq:velcor}
\end{equation}
Here, $m_i$ and ${\overrightarrow v}_i$ are the mass and the velocity of $i^{\rm th}$ particle, respectively.
$c_{\rm vv}(t)$ is calculated under the microcanonical ensemble with $\Delta t = 0.1$ fs and the data is sampled for 10 ps
with an output data frequency of $5 \times 10^{-4}$ ps. A representative $c_{\rm vv}(t)/c_{\rm vv}(0)$ 
for PLA is shown in Fig.~\ref{fig:vdosPLA}(a). The long lived fluctuations are clearly visible in global $c_{\rm vv}(t)$ 
that originates from the superposition of normal modes and thus its Fourier transform results in $g(\nu)$~\cite{Horbach1999JPCB,martin21prm}
using,
\begin{equation}
    g(\nu) = \frac {1}{C}\int_{0}^{\infty} \cos(2\pi \nu t) \frac{c_{\rm vv}(t)}{c_{\rm vv}(0)} {\rm d}t,
    \label{eq:vdos}
\end{equation}
where the pre--factor $C$ ensures $\int g(\nu) {\rm d}\nu = 1$. Fig.~\ref{fig:vdosPLA}(b) 
and Fig.~\ref{fig:gnu_4pol} show $g(\nu)$ for a PLA, and another set of four commodity polymer samples, respectively.
It can be appreciated that there are many high $\nu$ modes in these systems, i.e., for $\nu > \nu_{\rm room} \simeq 6.2$ THz,
that contribute rather non--trivially at a given $T$.

Given the discussions above, if the contribution of different modes are not properly accounted within a calculations, 
it will automatically lead to a wrong estimate of $\kappa$. Therefore, in the next sections, $g(\nu)$ 
shown in Figs.~\ref{fig:vdosPLA} and \ref{fig:gnu_4pol} will be used to accurately compute $\kappa$.

\subsection{Approach I: Computation of $\kappa$ using $g(\nu)$}
\label{ss:keq}

\begin{table*}[ptb]
\caption{A table listing the longitudinal ${v}_{\ell}$ and transverse ${v}_{t}$ sound velocities, 
the Debye frequency $\nu_{\rm D}$, the Debye temperature $\Theta_{\rm D}$, the experimental thermal transport 
         coefficient $\kappa^{\rm exp}$. The thermal transport coefficient $\kappa$ calculated using Eq.~\ref{eq:mtm} is listed together with
         its variation with respect to $\kappa^{\rm exp}$. The experimental values of ${v}_{\ell}$, ${v}_{t}$, and
         $\kappa^{\rm exp}$ are taken from Ref.~\cite{Cahill16Mac}, while the total atom number density $\rho_{\rm N}$ and 
         vibrational density of states $g(\nu)$ are computed from simulations. The data is shown for
         a temperature $T = 300$ K and for four different polymers, namely; poly(methyl methacrylate)(PMMA), poly(acrylic acid) (PAA), polyacrylamide (PAM), and poly(N-acryloyl piperidine) (PAP).}
\begin{center}
        \begin{tabular}{|c|c|c|c|c|c|c|c|c|c|c|c|}
\hline
Polymer  & ${v}_{\ell}^{\rm exp}$ (nm/ps)  & ${v}_{t}^{\rm exp}$ (nm/ps) & ~~$\nu_{\rm D}$ (THz)~~ & ~~$\Theta_{\rm D}$ (K)~~ &$\kappa^{\rm exp}$ (Wm$^{-1}$K$^{-1}$) & $\kappa$ (Wm$^{-1}$K$^{-1}$)
& $\frac {\left| \kappa^{\rm exp} - \kappa \right|} {\kappa^{\rm exp}}$ (\%)\\ 
\hline
\hline
\hline
PMMA     &  2.85 &  1.30 &  3.75 &   180.27     & 0.20 &   0.21& 5.0  \\
PAA      &  3.74 &  1.72 &  4.60 &   220.93     & 0.37 &   0.41& 10.8 \\
PAM      &  4.34 &  1.82 &  4.88 &   234.17     & 0.38 &   0.32& 15.8 \\
PAP      &  2.64 &  1.30 &  3.80 &   182.58     & 0.16 &   0.20& 25.0 \\
\hline 
\end{tabular}  \label{tab:kappa}
\end{center}
\end{table*}

The first approach is based within the framework of the well--known minimum thermal conductivity model (MTCM)~\cite{Cahill90PRB}.
To this end, the general expression of $\kappa$ for a 3--dimensional isotropic material reads~\cite{Kappabook},
\begin{equation}
    \kappa (T) = \left (\frac {\rho_{\rm N} h^2} {3k_{\rm B} T^2}\right)
    \sum_i \int \tau(\nu) v_{{\rm g},i}^2(\nu) 
    \frac {\nu^2 e^{{h\nu}/{k_{\rm B}T}}} {\left(e^{{h\nu}/{k_{\rm B}T}} -1 \right)^2}
    g(\nu) {\rm d}\nu,
    \label{eq:k_3d}
\end{equation}
where $\rho_{\rm N} = N/V(T)$ is the total atomic number density, $N$ the total number of atoms, 
and $h$ the Planck constant.
%
Starting with Eq.~\ref{eq:k_3d} and for the non--conducting amorphous solids, MTCM proposed that $\Lambda$ is limited to half 
the phonon wavelength and thus approximates $\tau(\nu) = 1/2\nu$~\cite{Cahill90PRB,Kappabook}. 
Also, $v_{\rm g} \simeq v_i$ with $v_i$ being the components of sound wave velocity. 
This description gives,
\begin{equation}
    \kappa = \left (\frac {\rho_{\rm N} h^2} {6k_{\rm B} T^2}\right)
    \left (v_{\ell}^2 + 2v_{t}^2 \right) {\mathcal I}(T),
    \label{eq:mtm}
\end{equation}
and
\begin{equation}
    {\mathcal I}(T) =  \int \frac {\nu e^{{h\nu}/{k_{\rm B}T}}} {\left(e^{{h\nu}/{k_{\rm B}T}} -1 \right)^2}
    g(\nu) {\rm d}\nu.
    \label{eq:mtm1}
\end{equation}
$v_{\ell} = \sqrt{C_{\rm 11}/\rho_{\rm m}}$ and $v_{t} = \sqrt{C_{\rm 44}/\rho_{\rm m}}$ are the longitudinal and the transverse sound wave velocities, respectively. 
Here, $C_{\rm 11} = K + 4 C_{\rm 44}/3$, $K$ is the bulk modulus, $C_{\rm 44}$ is the shear modulus, and $\rho_{\rm m}$ is the mass density.
It can be appreciated in Eq.~\ref{eq:mtm} that $\kappa$ is directly related to the materials stiffness via $v_{\ell}$ and $v_{t}$. 

Standard theoretical approaches typically use the Debye form of parabolic density of states $g_{\rm D}(\nu) = 3 \nu^2/\nu_{\rm D}^3$ in Eq.~\ref{eq:mtm},
where $\nu_{\rm D}$ is the Debye frequency~\cite{Horbach1999JPCB,Kappabook},
\begin{equation}
    \nu_{\rm D} = \left( \frac {9 \rho_{\rm N}} {4\pi} \right)^{1/3} \left( \frac {1} {v_{\ell}^3} + \frac {2}{v_{t}^3} \right)^{-1/3}.
\end{equation}
The Debye temperature $\Theta_{\rm D} = h \nu_{\rm D}/k_{\rm B}$. In Table~\ref{tab:kappa}, $\nu_{\rm D}$ and $\Theta_{\rm D}$
values are listed for four different polymers. Note that these values are calculated using the 
experimental data of $v_{\ell}$ and $v_{t}$ taken from Ref.~\cite{Cahill16Mac}, 
while $\rho_{\rm N}$ are from our simulations.
It can be seen that $\Theta_{\rm D}$ (or $\nu_{\rm D}$) are about 20--40\% lower than $T = 300$ K (or $\nu_{\rm room} = 6.2$ THz),
which is expected because of the dominant non--bonded interactions. 
Something that speaks in this favor is that $\Theta_{\rm D}$ for the weak vdW systems (PAP and PMMA) are 
about 40--50 K lower than the polymers dictated by a relatively stronger H--bonds (PAA and PAM).

The choice of $g_{\rm D}(\nu)$ is certainly a good approximation for the standard (non--polymeric) solids
where typically $\Theta_{\rm D} \gg T = 300$ K. However, $g_{\rm D}(\nu)$ in polymers (having rather complex 
$g(\nu)$, as in Fig.~\ref{fig:vdosPLA}(b) and Fig.~\ref{fig:gnu_4pol}) often tend to overestimate the contributions 
from the low frequency vibrational modes.
Something that supports this claim is that MTCM using $g_{\rm D}(\nu)$ predicts values comparable to $\kappa^{\rm exp}$ in
non--polymeric amorphous solids~\cite{Cahill90PRB}. However, for the polymers under the high temperature conditions, MTCM 
systematically estimates higher values than $\kappa^{\rm exp}$~\cite{Cahill16Mac}.

\begin{figure}[ptb]
\includegraphics[width=0.43\textwidth]{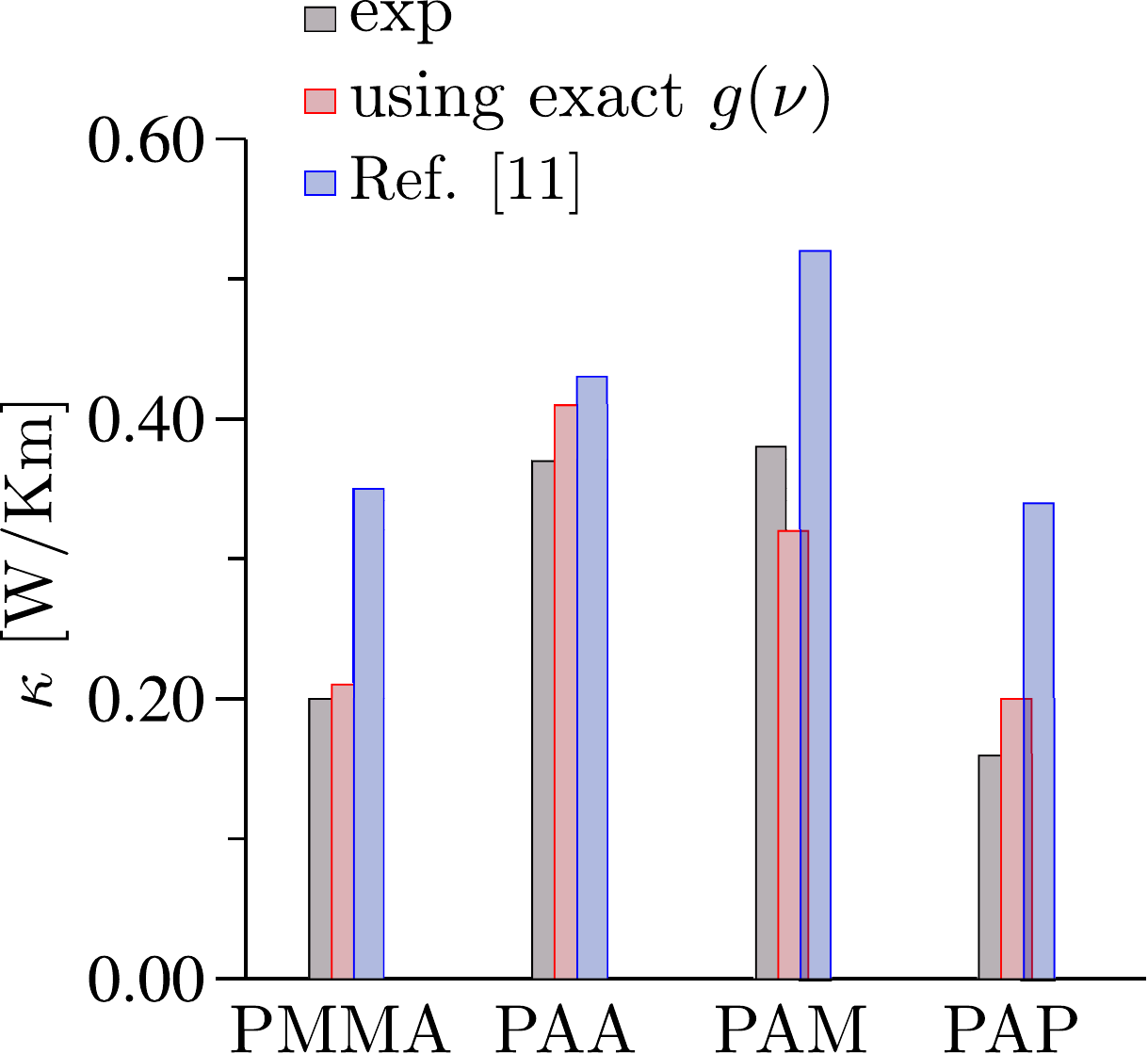}
\caption{A comparative plot of the thermal transport coefficient $\kappa$ at a temperature $T = 300$ K. The data is shown for four different polymers,
namely; poly(methyl methacrylate)(PMMA), poly(acrylic acid) (PAA), polyacrylamide (PAM), and poly(N-acryloyl piperidine) (PAP). 
The experimental data for these polymers are taken from Ref.~\cite{Cahill16Mac}, while the simulation data is 
calculated using the exact density of states $g(\nu)$ from Fig.~\ref{fig:gnu_4pol} in Eq.~\ref{eq:mtm}. 
For comparison, we have also included the high $T$ estimates from the minimum thermal conductivity~\cite{Cahill16Mac}.}
\label{fig:mtcm}
\end{figure}

This study revisits MTCM using the exact $g(\nu)$ from Fig.~\ref{fig:gnu_4pol} in Eq.~\ref{eq:mtm}.
Computed $\kappa$ for four different systems are listed in Table~\ref{tab:kappa}.
It can be appreciated that $\kappa$ matches within 1--25\% of $\kappa^{\rm exp}$.
An illustrative plot comparing $\kappa$ values between different approaches are compiled in Fig.~\ref{fig:mtcm}.

It should also be noted that the stiffness of a polymeric material is dictated by the non--bonded interactions that are classical in nature. 
Therefore, carefully conducted classical polymer simulations can give reasonable estimates of the elastic moduli 
comparable to the corresponding experimental values. On the contrary, however, quantum effects are important in 
the crystalline solids, i.e., for $T \ll \Theta_{\rm D}$~\cite{Martin01jcp,MHM01prb}. 
It might also be important to highlight that the $\tau(\nu) \propto 1/\nu$ behavior is valid for the amorphous systems 
under the high temperature conditions, while one may expect $\tau(\nu) \propto 1/\nu^2$ for the crystalline solids
strictly when the harmonic approximation holds~\cite{Feng2014}.

\subsection{Approach II: Scaling $\kappa$ using quantum estimate of heat capacity}

The results in Section~\ref{ss:keq} are presented at one thermodynamic state point, i.e., at $T = 300$ K.
This is specifically because, to the best of our knowledge, the data for $v_{\ell}$ and $v_{t}$ are not available
over a range of $T$ for the polymers listed in Table~\ref{tab:kappa}. Therefore, in this section, a slightly 
different (yet related) framework is used to compute $T-$dependent $\kappa(T)$. 
For this purpose, the classical estimate of the thermal transport coefficient $\kappa^{\rm cl}$ is first 
calculated using the approach--to--equilibrium (ATE) method~\cite{Lampin2013}.

\begin{figure*}[ptb]
\includegraphics[width=0.94\textwidth]{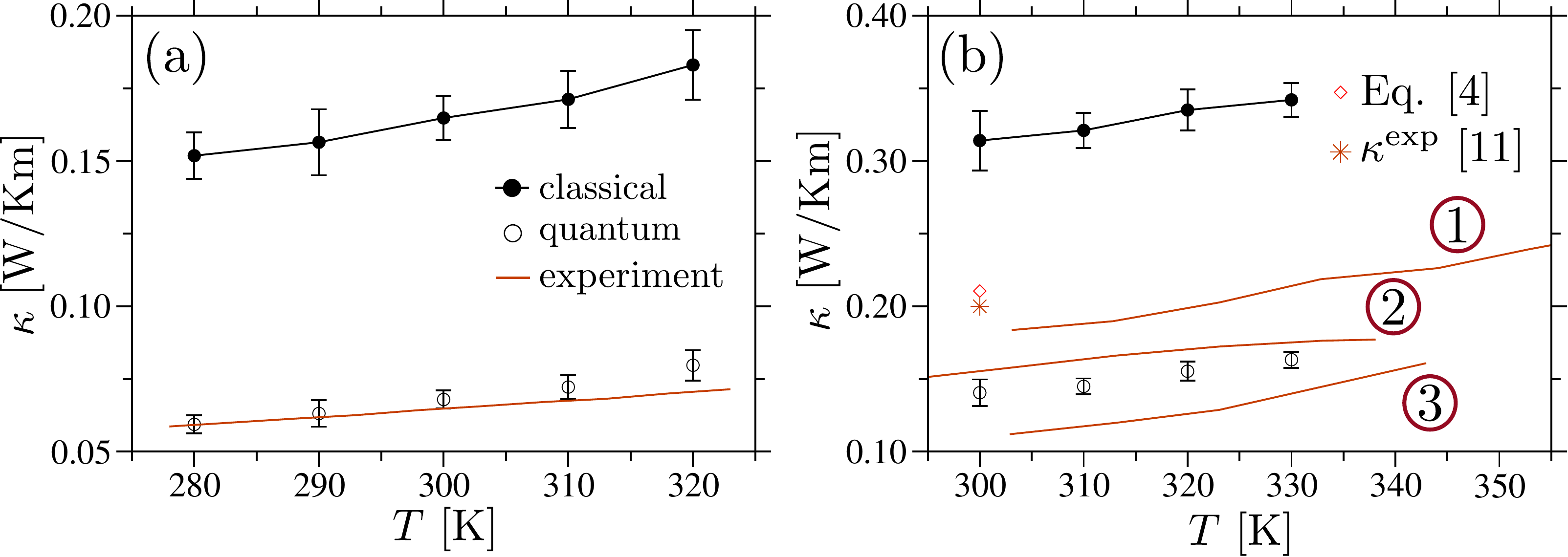}
\caption{Thermal transport coefficient $\kappa$ as a function of temprature $T$ for poly(lactic acid) (PLA) (part a) and poly(methyl methacrylate) (PMMA) (part b).
For comparison experimental data are also included. The experimental data of PLA is taken from Ref.~\cite{plakappa}.
The PMMA experimental data sets 1--3 are taken from Refs.~\cite{pmmalocalized,PMMACpexp,pmma14exp}, respectively.}
\label{fig:kappaT}
\end{figure*}

In ATE, a simulation box with a length $L_{x}$ along the $x-$direction is divided into three regions,
i.e., the middle region of width $L_x/2$ is sandwiched between two side regions of equal width $L_x/4$. 
The middle slab is kept at an elevated temperature $T_{\rm Hot} = T + 50$ K, while the two side slabs are maintained at 
a lower temperature $T_{\rm Cold} = T - 50$ K. Here, $T_{\rm Hot}$ and $T_{\rm Cold}$ are the kinetic temperatures of 
the hot and the cold regions, respectively. $T$ refers to a reference temperature at which $\kappa^{\rm cl}(T)$ is calculated. 
In the first step, these regions are thermalized under the canonical simulations for 5 ns with $\Delta t = 1$ fs. 
After this stage, $\Delta T(t) = T_{\rm Hot}-T_{\rm Cold}$ is allowed to relax during a set of microcanonical 
runs for 50 ps with $\Delta t = 0.1$ fs. 
From an exponential relaxation of $\Delta T(t) \propto \exp \left(-t/\tau_x\right)$, the
time constant $\tau_x$ for the energy flow along the $x-$direction is calculated. Finally, $\kappa^{\rm cl}(T)$ 
can be estimated using~\cite{Lampin2013},
\begin{equation}
\label{eq:ate}	
	\kappa^{\rm cl} (T) = \frac {1}{4\pi^2} \frac {c^{\rm cl}(T) L_x} {A \tau_x}.
\end{equation}
$A$ is the cross--section area of a sample. The classical estimate of specific heat is calculated using
$c^{\rm cl}(T) =  {H(T + \Delta T) - H(T -\Delta T)}/{2 \Delta T}$ and the enthalpy is $H(T) = U(T) + pV(T)$.
$U(T)$ is the internal energy including the mean kinetic energy, $p = 1$ atm is the external pressure, and 
$V(T)$ is the system volume. Computed $\kappa^{\rm cl}(T)$ for the PLA and PMMA samples are shown in Fig.~\ref{fig:kappaT}, 
see the $\bullet$ data sets. It is clearly visible that $\kappa^{\rm cl}(T)$ is about a factor of 2--to--3 times larger than the 
corresponding $\kappa^{\rm exp}(T)$.

Note that for $T > \Theta_{\rm D}$ (as in the commodity polymers listed in Table~\ref{tab:kappa}), 
$\tau_x$ is dominated by the non--bonded (classical) interactions~\cite{pmmalocalized}.
On the contrary, the intra--molecular stiff interactions along a chain contour do not contribute to $\tau_x$, 
yet they are by default incorporated in $c^{\rm cl}(T)$ and thus $\kappa^{\rm cl}(T) > \kappa^{\rm exp}(T)$.
This is consistent with the data that the difference between $c^{\rm cl}(T)$ and $c^{\rm exp}(T)$ is also about 
a factor of 2--to--3, see the $\bullet$ data sets and the lines in the Appendix Fig.~\ref{fig:CpT}.

Within the above discussion, if one can use the accurate estimate of $c(T)$ in Eq.~\ref{eq:ate} by properly accounting for the 
contributions from the vibrational mode at a given $T$~\cite{martin21prm,Zaccone21pre}, one may just simply get the quantum corrected estimate of $\kappa(T)$. 
For this purpose, $c(T)$ is calculated using a recently proposed method~\cite{martin21prm}. 
In a nutshell, this method uses the Binder approach~\cite{Horbach1999JPCB} to estimate the contributions of the stiff harmonic modes
and thus their total contribution is given by,
\begin{equation}\label{eq:ccor}
    {\frac {c_{\rm h}(T)} {k_{\rm B}}} = \frac {h^2}{k_\textrm{B}^2 T^2} \int_{0}^{\infty} \frac {\nu^2 e^{{h\nu}/{k_{\rm B}T}}} {\left(e^{{h\nu}/{k_{\rm B}T}} -1 \right)^2}
    g(\nu) {\rm d}\nu,
\end{equation}
which is then used to get the difference $\Delta c(T)$ between the classical and the quantum descriptions,
\begin{equation} \label{eq:chrel}
    \frac {\Delta c(T)} {k_{\rm B}} = \int_0^\infty \left\{ 1 - 
    \left(\frac {h\nu}{k_\textrm{B}T}\right)^2 \frac {e^{{h\nu}/{k_{\rm B}T}}} {\left(e^{{h\nu}/{k_{\rm B}T}} -1 \right)^2} \right\} g(\nu) {\rm d}{\nu},
\end{equation}
and finally gives the quantum corrected estimate~\cite{martin21prm},
\begin{equation}
    c(T) = c^{\rm cl}(T) - \Delta c(T).
    \label{eq:cpquant}
\end{equation}
The main advantage of this approach is that the contributions of the stiff harmonic modes are corrected, 
while the contributions from the soft (anharmonic) modes remain unaltered and thus does not alter the 
macroscopic polymer properties. In the Appendix Fig.~\ref{fig:CpT}, quantum corrected $c(T)$ for PLA and PMMA samples are shown. 
As expected, the quantum correction discussed above gives reasonably estimates of $c(T)$ in comparison to $c^{\rm exp}(T)$. 
The calculated $c(T)$ is then used to obtain quantum corrected $\kappa(T)$ using, 
\begin{equation}
    	\kappa(T) = \frac {1}{4\pi^2} \frac {c(T) L_x} {A \tau_x}.
    \label{eq:kcor1}
\end{equation}
The resultant data is shown in Fig.~\ref{fig:kappaT}, see the $\circ$ data sets. 
It can be appreciated that this simple approach in Eq.~\ref{eq:kcor1} gives reasonable estimates of $\kappa(T)$.

\section{Conclusions and discussions}
\label{sec:conc}

This work used a conventional classical molecular dynamics setup to estimate the quantum corrected thermal transport 
coefficient $\kappa$ in polymeric solids. For this purpose, the exact vibrational density of states $g(\nu)$ 
is used as a key observable within two different (yet related) semi--analytical approaches. 
In one approach, $\kappa$ is computed within the framework of the minimum thermal conductivity model~\cite{Cahill90PRB}, 
while another approach simply uses the quantum estimate of specific heat $c(T)$ as a correction to $\kappa(T)$. 
The data for a set of five different commodity polymers show reasonable agreement with the experiments,
at one thermodynamic state point and also with changing $T$. Therefore, this work attempts to highlight
a couple of simple approaches to obtain quantum $\kappa$ from classical simulations.
The approaches presented herein can also be used in studying $\kappa$ under the high pressure conditions.
One key application is in the field of hydrocarbon--based oils~\cite{kappaOil,martin21prm} under high pressures.

It is also important to highlight that the approaches discussed here is valid for the non--conducting amorphous polymers,
where localized vibrations carry the heat current~\cite{pmmalocalized}. These vibrations are dictated by the non--bonded interactions 
between the neighboring monomers and are classical in nature. Therefore, by simply eliminating the contributions of the 
intra--molecular stiff modes give reasonable estimates of $\kappa(T)$. However, when dealing with the chain oriented systems~\cite{shen2010polyethylene}, 
such as in the polymer fibers~\cite{shen2010polyethylene} or in the molecular forests~\cite{bhardwaj2021thermal}, the situation 
is somewhat different. 
This is particularly because $\kappa$ is an extended configuration is dominated by the stiff intra--molecular interactions,
i.e., almost a representative of the crystalline structures along the chain backbone~\cite{crystPol2,crystPol1,mukherji24lang}.
For example, a standard amorphous polymer usually has $\kappa \simeq 0.1-0.4$ W/Km~\cite{Pipe15NMat,Cahill16Mac,Keblinski20,James22CMS},
while the expanded chain configurations usually have $\kappa \ge 100$ W/Km~\cite{shen2010polyethylene,bhardwaj2021thermal}.
Another set of systems where intra--molecular interactions dictate $\kappa$ is the highly cross--linked networks, 
where a delicate balance between between the bond density, network micro--structure, and 
bond property controls $\kappa$~\cite{cahill21acsappm,DMPRM21,lv2021effect}.

A simplistic scaling correction in Eq.~\ref{eq:kcor1} may not be appropriate for the crystalline solids with long range order, 
where propagating phonons carry the heat current~\cite{Kappabook}. In this context, it was readily observed that the representative 
hump in $\kappa^{\rm exp}(T)$ for the crystals happen at a $T$ that is far lower than the typical plateau in $c(T)$,
i.e., the anharmonic effects already become relevant at a $T \ll \Theta_{\rm D}$. 
For example, in crystalline silicon, a hump in $\kappa$ happens between 10--20 K~\cite{CrystalEXP}, while $\Theta_{\rm D} > 600$ K.
In such systems, therefore, quantum effects must be properly incorporated via $\tau(\nu,T)$, $v_{\rm i}(T)$, and also $c(T)$ in Eq.~\ref{eq:mtm}.\\

\noindent{\bf Acknowledgement:}
The content presented in this work would not have been possible without numerous very stimulating discussions with Martin M\"user
during the preparation and after the publication of our collaborative work in Ref.~\cite{martin21prm}, whom I take this opportunity to gratefully acknowledge.
I also thank Robinson Cortes--Huerto for the very useful discussions.
I further thank the Advanced Research Computing facility where simulations are performed.
This research was undertaken thanks, in part, to the Canada First Research Excellence Fund (CFREF), Quantum Materials and Future Technologies Program.

\appendix

\section{Quantum Heat Capacity of polymers}
\label{app:A}

\begin{figure}[ptb]
\includegraphics[width=0.49\textwidth]{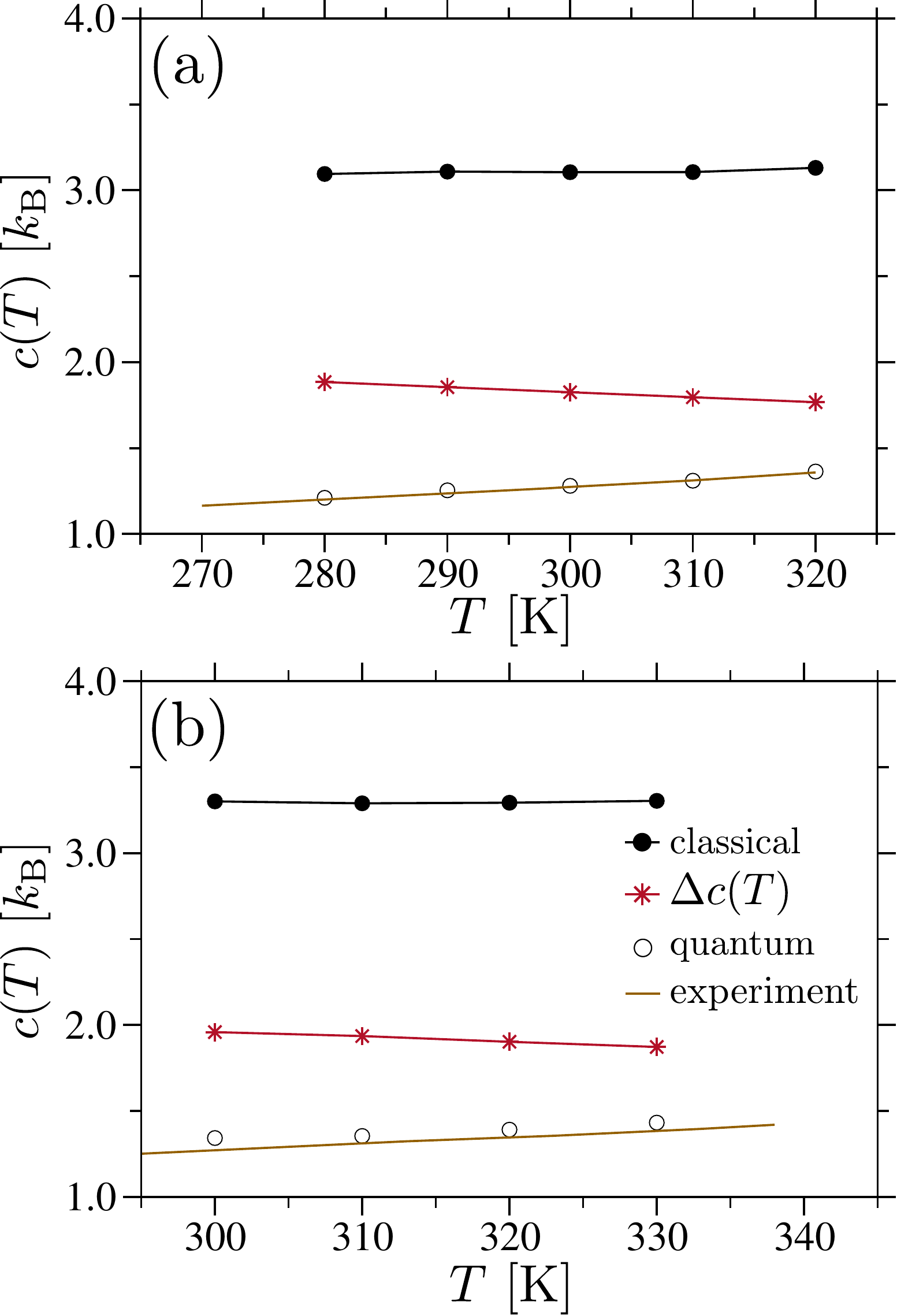}
\caption{Per particle specific heat $c(T)$ as a function of temperature $T$ for poly(lactic acid) (PLA) (part a) and poly(methyl methacrylate) (PMMA) (part b).
The data is shown for the classical estimate of specific heat $c^{\rm cl}(T)$, correction factor $\Delta c(T)$ using Eq.~\ref{eq:chrel}, and quantum specific heat $c(T)$ using Eq.~\ref{eq:cpquant}. 
For comparison experimental data are also included. The PLA and PMMA data is taken from Ref.~\cite{PLACpexp} and from Ref.~\cite{PMMACpexp}, respectively.}
\label{fig:CpT}
\end{figure}

Fig.~\ref{fig:CpT} shows the comparative data sets for heat capacity. 
While classical estimate $c^{\rm cl}(T)$ is overestimated, simple quantum correction using Eq.~\ref{eq:cpquant} eliminates 
the unwanted contributions from the high $\nu$ stiff modes and thus gives a better comparison between $c(T)$ and $c^{\rm exp}(T)$.


\end{document}